\documentclass[conference]{IEEEtran}
\IEEEoverridecommandlockouts
\usepackage{booktabs}
\usepackage{cite}
\usepackage{subcaption}
\usepackage[margin=1in]{geometry}
\usepackage{amsmath,amssymb,amsfonts}
\usepackage{algorithmic}
\usepackage{graphicx}
\usepackage{textcomp}
\usepackage{xcolor}
\usepackage{array}
\usepackage{hyperref}
\usepackage{balance}

\def\BibTeX{{\rm B\kern-.05em{\sc i\kern-.025em b}\kern-.08em
    T\kern-.1667em\lower.7ex\hbox{E}\kern-.125emX}}

\usepackage{mdframed}

\def\BibTeX{{\rm B\kern-.05em{\sc i\kern-.025em b}\kern-.08em
    T\kern-.1667em\lower.7ex\hbox{E}\kern-.125emX}}

\renewcommand{\footnoterule}{%
  \kern -3pt                        
  \hrule width 0.4\columnwidth      
  \kern 2.6pt                       
}
\def\BibTeX{{\rm B\kern-.05em{\sc i\kern-.025em b}\kern-.08em
    T\kern-.1667em\lower.7ex\hbox{E}\kern-.125emX}}

\newmdenv[
    backgroundcolor=gray!10,
    linecolor=gray!10,      
    shadow=true,
    shadowsize=2pt,
    shadowcolor=black!50,
    innertopmargin=5pt,
    innerbottommargin=5pt,
    innerleftmargin=5pt,
    innerrightmargin=5pt,
    linewidth=0pt,          
]{keytakeaway}

\usepackage{hyperref}

\makeatletter
\def\ps@IEEEtitlepagestyle{%
  \def\@oddfoot{\hbox{}\scriptsize
    \hfil
    \parbox{0.95\textwidth}{\centering
      \textcopyright~2025 IEEE. Personal use of this material is permitted.
      Permission from IEEE must be obtained for all other uses, in any current or future media,
      including reprinting/republishing this material for advertising or promotional purposes,
      creating new collective works, for resale or redistribution to servers or lists, or reuse
      of any copyrighted component of this work in other works.\\[2pt]
      DOI: \href{https://doi.org/10.1109/SOSE67019.2025.00006}{10.1109/SOSE67019.2025.00006}
    }
    \hfil}%
  \def\@evenfoot{}%
}
\makeatother

\begin{document}

\title{Silent Failures in Stateless Systems: Rethinking Anomaly Detection for Serverless Computing
\\
}



\author{
    \IEEEauthorblockN{Chanh Nguyen, Erik Elmroth, and Monowar Bhuyan}
    \IEEEauthorblockA{\textit{Department of Computing Science, Ume{\aa} University}\\ 
    \textit{SE-90187 Ume{\aa}, Sweden}\\
    Emails: \{chanh, elmroth, monowar\}@cs.umu.se} 
}

\maketitle

\begin{abstract}
Serverless computing has redefined cloud application deployment by abstracting infrastructure and enabling on-demand, event-driven execution, thereby enhancing developer agility and scalability. However, maintaining consistent application performance in serverless environments remains a significant challenge. The dynamic and transient nature of serverless functions makes it difficult to distinguish between benign and anomalous behavior, which in turn undermines the effectiveness of traditional anomaly detection methods. These conventional approaches, designed for stateful and long-running services, struggle in serverless settings where executions are short-lived, functions are isolated, and observability is limited. 

In this first comprehensive vision paper on anomaly detection for serverless systems, we systematically explore the unique challenges posed by this paradigm, including the absence of persistent state, inconsistent monitoring granularity, and the difficulty of correlating behaviors across distributed functions. We further examine a range of threats that manifest as anomalies, from classical Denial-of-Service (DoS) attacks to serverless-specific threats such as Denial-of-Wallet (DoW) and cold start amplification. 
Building on these observations, we articulate a research agenda for next-generation detection frameworks that address the need for context-aware, multi-source data fusion, real-time, lightweight, privacy-preserving, and edge-cloud adaptive capabilities.

Through the identification of key research directions and design principles, we aim to lay the foundation for the next generation of anomaly detection in cloud-native, serverless ecosystems.



\end{abstract}


\section{Introduction}

Over the past decade, serverless computing has emerged as a defining paradigm in cloud-native application development~\cite{aslanpour2021serverless, li2022serverless}. 
While traditional Infrastructure-as-a-Service (IaaS) and Platform-as-a-Service (PaaS) models offer varying degrees of automation in provisioning and scaling, serverless platforms go further by providing per-request automatic provisioning, transparent scaling, and fine-grained billing based on actual execution time~\cite{jonas2019cloud, rajan2018serverless}. 
These advantages eliminate the need for developers to manage runtime environments, instance lifecycles, or idle resource allocation. 
As a result, serverless computing has seen widespread adoption across a range of domains, including web services~\cite{shafiei2022serverless}, machine learning workflows~\cite{carreira2019cirrus,yang2022infless, hui2025exploring}, IoT backends~\cite{merlino2024faas}, and high-throughput data processing pipelines~\cite{werner2024reference, nastic2017serverless}.
This adoption has been facilitated by the development of a diverse ecosystem of serverless platforms,  such as commercial offerings  AWS Lambda\footnote{\url{https://aws.amazon.com/lambda/}}, Google Cloud Functions\footnote{\url{https://cloud.google.com/functions}}, and Azure Functions\footnote{\url{https://azure.microsoft.com/en-us/services/functions}}, as well as open-source frameworks like Apache OpenWhisk\footnote{\url{https://openwhisk.apache.org/}} and OpenFaaS\footnote{\url{https://www.openfaas.com/}}, which offer developers increased flexibility in deployment and greater operational control.

While serverless platforms offer simplified deployment and automatic scalability, they also fundamentally reshape the operational model of cloud applications. 
First, the ephemeral and stateless nature of the execution model introduces unique observability challenges.
Empirical studies show that the majority of serverless functions are short-lived, often completing within milliseconds to a few seconds~\cite{bauer2024globus, shahrad2020serverless, joosen2025serverless}. This temporal brevity presents a fundamental mismatch with traditional monitoring tools, which are designed for long-running, stateful services.
Consequently, collecting sufficient runtime telemetry for profiling or debugging becomes a significant challenge.
Second, unlike traditional virtual machine (VM) or container-based deployments, where developers can instrument the infrastructure and access detailed system telemetry (i.e., CPU utilization, memory usage, queue delays, and container lifecycle events), serverless platforms abstract away the execution environment entirely. This lack of visibility limits their ability to diagnose anomalies or understand performance regressions, especially in latency-sensitive or cost-critical applications~\cite{sharma2023challenges, li2022securing, raza2021sok}. 
Third, from perspective of insfrastructure providers, serverless computing shifts much of the operational responsibility to the provider, who must dynamically schedule and scale ephemeral function instances, manage cold start pools, enforce tenant isolation, and maintain system responsiveness under highly bursty and unpredictable workloads~\cite{ascigil2021resource}. This architectural shift makes system-level anomaly detection significantly more challenging: anomalies may manifest as degraded scheduling efficiency, cold start thrashing, unfair resource contention, or orchestration delays, often without clear or persistent symptoms. 
In distributed, heterogeneous server deployments, these challenges are further exacerbated by resource constraints, architectural diversity, and limited global visibility across nodes, making robust and timely detection of system anomalies even more difficult~\cite{ascigil2021resource, russo2023serverledge}.


In this paper, we take a step back to examine the unique challenges and open questions surrounding anomaly detection in serverless platforms. Rather than proposing a single detection mechanism, we present a systematic exploration of the fundamental constraints that make anomaly detection in this domain uniquely difficult.  We analyze potential sources of anomalies and attack surfaces, including cold start misbehavior, control-plane bottlenecks, unpredictable scheduling latencies, and workload-induced contention. Building on these observations, we articulate a vision for rethinking observability and anomaly detection in serverless systems, one that emphasizes request-level introspection, distributed event correlation, and adaptive telemetry. Our goal is to illuminate a space of design opportunities and research directions toward robust, scalable, and practical anomaly detection solutions across the spectrum of serverless computing, from centralized cloud environments to distributed, resource-constrained edge deployments.

\textit{Organization}. The remainder of this paper is organized as follows. In Section~\ref{sec:bg_mt}, we present background and motivation, highlighting the limitations of existing monitoring and anomaly detection approaches in the context of serverless systems. In Section~\ref{sec:threat}, we characterize the landscape of anomalies and attack surfaces specific to serverless platforms. In Section~\ref{sec:vision}, we outline a forward-looking vision for anomaly detection in serverless computing, identifying key research directions and design considerations. Section~\ref{sec:opp} presents emerging opportunities and outlines several promising directions for future investigation. Section~\ref{sec:challenges} examines the practical obstacles to realizing this vision in real-world deployments, including telemetry limitations, data scarcity, generalization barriers, deployment overhead, and evolving threat models. 
Finally, Section~\ref{sec:conclus} concludes the paper by summarizing our key insights, reiterating our vision for next-generation anomaly detection, and outlining its broader implications for robust serverless ecosystems.

\section{Background and Motivation}
\label{sec:bg_mt}

\begin{figure*}[htbp]
  \centering
  \begin{subfigure}[b]{0.45\linewidth}
    \includegraphics[width=\linewidth]{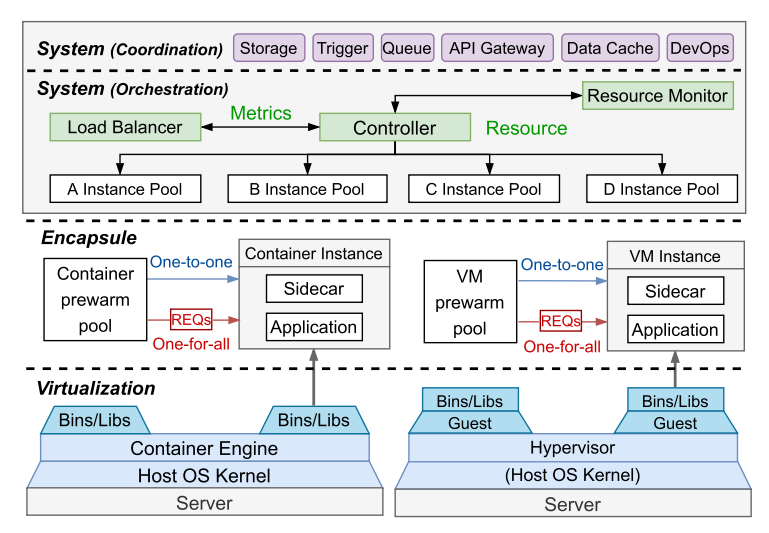}
    \caption{Classical four-layer serverless architecture, source~\cite{li2022serverless}.}
    \label{fig:plotA}
  \end{subfigure}
  \hfill
  \begin{subfigure}[b]{0.4\linewidth}
    \includegraphics[width=\linewidth]{graph/OpenWhisk_architecture.png}
    \caption{Apache OpenWhisk architecture, source~\cite{openwhisk}.}
    \label{fig:plotB}
  \end{subfigure}
  \caption{Overview of serverless system design: (a) classical layered architecture, and (b) an implementation example with Apache OpenWhisk.}
  \label{fig:twoPlots}
\end{figure*}

\noindent \textbf{Serverless computing} represents a fundamental shift in how cloud applications are built and operated. By abstracting away server management, resource provisioning, and scaling, serverless platforms such as AWS Lambda, Google Cloud Functions, and Azure Functions allow developers to focus exclusively on writing event-driven functions. 
To comprehend the operational mechanisms of a serverless platform, we first examine its architecture and the typical control flow during function invocation. Figure~\ref{fig:plotA} illustrates a generalized four-layer architecture. Figure~\ref{fig:plotB} then illustrates how this architecture is concretely realized within Apache OpenWhisk. 

In general, the 4-layer architecture of a serverless platform comprises the following components:

\begin{itemize}
    \item \textbf{Coordination layer:} Handles external interactions and exposes endpoints (e.g., REST APIs, SDKs) through which users deploy and invoke functions.
    
    \item \textbf{Orchestration layer:} Responsible for request routing, scheduling, and load balancing. This layer includes components such as the controller in OpenWhisk, which manages invocation dispatching.
    
    \item \textbf{Encapsulation layer:} Manages compute resources, typically a pool of containers or lightweight VMs, that execute functions in isolated environments.
    
    \item \textbf{Infrastructure layer:} Supports low-level resource provisioning, network virtualization, and storage services, typically abstracted by container runtimes and orchestration platforms such as Kubernetes.
\end{itemize}

When a function invocation is received, it first passes through the coordination layer, which includes the API gateway (e.g., Nginx\footnote{\url{https://nginx.org/}} in OpenWhisk). 
The request is then forwarded to the \textit{controller} in the orchestration layer, which temporarily buffers the invocation and attempts to schedule it onto an available runtime  in the encapsulation layer (e.g., an invoker in OpenWhisk), responsible for provisioning or reusing a containerized environment to execute the function.

In serverless computing, a \textit{warm container} refers to a pre-initialized container ready to handle incoming requests immediately. In contrast, a \textit{cold start} occurs when no warm container is available, requiring the system to provision a new instance and initialize the runtime and dependencies. This introduces non-negligible latency. 
Prior studies~\cite{xiao2024making, bauer2024empirical} report that cold start delays on platforms such as AWS Lambda and Microsoft Azure can exceed execution times by 16$\times$ to 166$\times$. To reduce this overhead, providers often rely on static keep-alive windows (e.g., 10--20 minutes)~\cite{shahrad2020serverless}, or adopt prewarming techniques that proactively initialize containers with common runtimes and libraries based on anticipated load.

To ensure efficient operation, serverless platforms also enforce runtime control mechanisms.  Each function invocation is subject to a \textit{timeout limit}, which bounds its maximum execution duration to prevent indefinite resource consumption (e.g., Amazon lambda max timeout limit 15 mins). Additionally, most platforms apply \textit{rate limits}, capping the number of concurrent invocations per user account or region to ensure fair resource allocation and system stability.

Analyses of real-world function traces~\cite{joosen2025serverless, bauer2024globus} demonstrate that many serverless functions execute within just a few milliseconds or seconds (notably, 85\% of all tasks complete in under one minute~\cite{bauer2024empirical}), and their invocation patterns can be highly sporadic, i.e., exhibiting bursts of activity within a single minute followed by extended idle periods lasting several minutes. This temporal volatility reflects the inherently short-lived and invocation-driven nature of serverless workloads, a behavior we refer to as \textit{ephemeral}~\cite{joosen2025serverless}. 
Such behavior introduces unique challenges for traditional anomaly detection methods, as discussed below.

\noindent\textbf{Traditional anomaly detection in cloud systems.} In conventional cloud environments, applications are deployed on long-running virtual machines (VMs) or containers, which persist over time and expose a rich set of resource and performance metrics, such as CPU utilization, memory usage, disk I/O, and network throughput. These metrics are typically collected by monitoring agents, sidecar containers, or telemetry tools such as Prometheus\footnote{\url{https://prometheus.io/}}, cAdvisor\footnote{\url{https://github.com/google/cadvisor}}, or Datadog\footnote{\url{https://www.datadoghq.com/}}. For instance, Prometheus periodically scrapes the \textit{/metrics} endpoint (e.g., every 15 seconds) and stores the results in a time-series database for subsequent analysis.

Anomaly detection in such environments often relies on statistical thresholding, time-series modeling, or supervised machine learning techniques trained on historical telemetry data~\cite{vallis2014novel, huang2017time, sauvanaud2018anomaly, garg2019hybrid, Deka2023}. These approaches assume stable and continuous execution environments, consistent data collection intervals, and persistent access to system-level signals.

While observability tools such as Prometheus can be configured to scrape metrics from short-lived services, doing so reliably at serverless timescales remains challenging. Serverless platforms abstract away the infrastructure layer, offering no direct access to node-level metrics or runtime context. Even when metrics or logs are made available (e.g., via AWS CloudWatch), they are typically delayed, coarse-grained, and insufficient for fine-grained anomaly detection in systems. Consequently, traditional anomaly detection methods struggle in serverless environments, where execution is transient, metric availability is sparse, and system-level visibility is limited.

In addition, serverless workloads are highly bursty and non-stationary~\cite{joosen2025serverless}, with functions triggered by external events that can cause abrupt spikes in invocation rates. Cold starts introduce further variability in response times and are often difficult to distinguish from genuine performance anomalies. Since scheduling, placement, and resource reuse are entirely managed by the platform, developers lack both visibility into and control over the underlying orchestration mechanisms. These constraints hinder the applicability of traditional detection signals, such as CPU saturation or memory leaks, which may be absent or unobservable in a serverless context.

\noindent\textbf{Motivation for rethinking anomaly detection.} The unique characteristics of serverless computing motivate a fundamental rethinking of how anomalies should be detected and diagnosed. From the perspective of serverless application developers or providers (i.e., users who deploy functions on a serverless platform), detection must shift from infrastructure-centric monitoring to a more \textit{request-centric} inference approach. Since direct observations of containers or hosts are typically hidden and only accessible to platform operators, detection systems must instead rely on behavioral signatures -- such as function invocation patterns (e.g., function A triggered after function B), response time fluctuations, cold start frequency, and inter-function delays. This calls for lightweight, data-efficient, and \textit{context-aware} techniques capable of extracting meaningful insights from fragmented telemetry.

From the perspective of the platform operator, or in custom platforms where deeper instrumentation is possible (e.g., OpenWhisk, OpenFaaS), access to low-level resource metrics (e.g., CPU, memory, I/O) alongside unstructured logs from internal components (e.g., controller traces, invoker logs, platform events) enables a more fine-grained view of system behavior. This opens the opportunity to develop log-data fusion mechanisms that correlate function-level behaviors with underlying infrastructure dynamics, thereby improving the accuracy and timeliness of anomaly detection.

In addition, the distributed and event-driven nature of serverless workloads, especially in cloud–edge deployments, introduces significant complexity for anomaly detection. Functions may execute across heterogeneous nodes with varying performance characteristics, network conditions, and geographic locations. Moreover, anomalies can arise not just from isolated failures but from cascading issues across microservices, message queues, or third-party APIs (e.g., Mailgun API\footnote{\url{https://documentation.mailgun.com/}}). In such environments, centralized detection approaches are often infeasible due to high latency, limited visibility, and scalability concerns. These realities call for detection mechanisms that are decentralized, capable of operating close to the execution point, while remaining low-latency and adaptive to dynamic execution contexts.

In essence, the operational realities of serverless computing present a profound architectural mismatch for conventional anomaly detection. This disconnect underscores the urgent need for a new generation of detection methodologies, ones inherently designed for ephemeral, opaque, and highly dynamic environments. Such methods must bridge the visibility gaps for developers and provide robust, fine-grained insights for operators. In the following sections, we characterize the unique challenges and emerging anomaly patterns in this space, and present a vision for developing more suitable, scalable, and adaptive detection techniques aligned with the serverless paradigm.

\section{Threat Landscape in Serverless Computing}
\label{sec:threat}
Anomalies, defined as observations deviating significantly from expected behavior and potentially indicating underlying issues~\cite{hawkins1980identification}, can manifest in various forms within serverless systems. These include \textit{point anomalies} (e.g., a single invocation exhibiting abnormal latency), \textit{collective anomalies} (e.g., a sequence of degraded executions), or \textit{contextual anomalies} (e.g., failures that occur only under specific workload conditions, such as peak traffic) \cite{Bhuyan2014}. In serverless environments, such anomalies commonly appear as increased cold start latency, reduced throughput, elevated error rates, or resource exhaustion, leading to degraded quality of service (QoS), violations of service-level agreements (SLAs), denial of service, and, in some cases, system-wide interruptions or outages. 

The architectural principles of serverless computing significantly reshape traditional system vulnerabilities and threat models~\cite{marin2022serverless, shen2022gringotts}. We categorize the key threats and their manifestations into two primary areas: (i) \textit{operational vulnerabilities}, which arise from the inherent dynamics of serverless resource management and affect availability and performance, and (ii) \textit{adversarial threats}, which originate from malicious actors exploiting these dynamics for financial, security, or service disruption goals.

\subsection{Operational vulnerabilities: internal stressors to availability and performance}
These threats primarily arise from the inherent characteristics of serverless platforms, such as unpredictable execution, resource management complexities, and dependencies on shared backend services. Anomalies in this domain typically reflect system inefficiencies or degradation.
\begin{itemize}
    \item \textbf{Cold start latency amplification:} The overhead of initializing new function instances (cold starts) can severely impact perceived performance and violate Quality of Service (QoS) guarantees.  Under bursty workloads, a surge of concurrent cold starts may amplify latency across the system, especially for latency-sensitive functions or edge deployments~\cite{vahidinia2020cold, ebrahimi2024cold}.
    \item \textbf{Resource contention and noisy neighbor effects:} Despite containerization and runtime isolation, serverless functions often share underlying compute infrastructure (e.g., CPU, memory, I/O). Under high load or on constrained edge nodes, this can lead to resource contention among co-located tenants, resulting in throughput degradation and performance variability~\cite{suresh2021servermore, agache2020firecracker}, commonly referred to as the ``noisy neighbor'' effect.
    \item \textbf{Orchestration delay and queuing collapse:} Serverless platforms rely on orchestration layers (e.g., controllers, load balancers, and schedulers) to manage invocation routing and instance lifecycle~\cite{ascigil2021resource}. Centralized bottlenecks or inefficient scheduling decisions, particularly under diurnal or regionally correlated bursts, can introduce significant dispatch latency. In extreme cases, this may trigger queuing collapse, cascading timeouts, or systemic unresponsiveness.
    \item \textbf{Backend and dependency failures:} Serverless functions frequently depend on external services, such as storage, message brokers, databases, or third-party APIs. Failures or degraded performance in these dependencies propagate to the function layer, manifesting as elevated error rates or long-tail latencies. Due to the decoupled nature of serverless architectures, identifying root causes across service boundaries remains a persistent challenge.
\end{itemize}

\subsection{Adversarial threats: external exploitation of serverless weaknesses}
Beyond system-internal sources of performance degradation, serverless platforms are increasingly vulnerable to adversarial threats. Malicious actors exploit the architectural characteristics and operational dynamics of serverless systems to induce service disruption, escalate costs, or evade detection. These threats often mirror or amplify benign anomalies, blurring the boundary between fault and attack. As a result, anomaly detection mechanisms must account for both accidental failures and deliberate manipulation of the control and data planes.

\begin{itemize}
    \item \textbf{Denial-of-Service (DoS) and Denial-of-wallet~(DoW) attacks:} While serverless platforms auto-scale to absorb traffic, a sustained high volume of requests targeting publicly exposed functions (e.g., via API Gateways) can still saturate configured concurrency limits~\cite{xiong2024warmonger}. This can lead to legitimate requests being queued or dropped, effectively denying service to legitimate users, especially if cold start overheads prevent rapid scaling. More subtly, DoW attacks~\cite{kelly2021denial, shen2022gringotts} aim to silently drain tenant budgets by repeatedly invoking costly functions, e.g., inference for machine learning (ML) models or data-intensive workflows. Even low-rate, persistent invocations can accumulate substantial cost, succeeding without triggering rate limits or violating correctness constraints.
    \item \textbf{Trigger abuse and event injection:} Serverless functions, by their event-driven nature, rely on diverse activation triggers like HTTP requests, object storage events, and message queues. Adversaries can exploit misconfigured, publicly exposed, or weakly authenticated triggers to invoke functions out of context or with crafted payloads~\cite{marin2022serverless}. For example, a malicious file uploaded to a monitored storage bucket could trigger sensitive data processing, or spoofed webhooks/timers might initiate unintended execution. When functions are chained in workflows (e.g., AWS Step Functions), such trigger abuse can cascade, amplifying the attack's impact across dependent services.


    \item \textbf{Evasion and obfuscation tactics:} The ephemeral and stateless nature of serverless functions can be exploited by attackers to evade detection and forensic analysis. Malicious logic can be segmented across multiple invocations or injected via environment variables, making it harder to identify a complete attack signature in any single event~\cite{ginzburg2020serverless,gopireddy2018behavioural}.  Due to asynchronous logging pipelines or sampling in highly concurrent environments, rapid-invocation attacks or quick bursts of malicious activity might escape full visibility in the platform's telemetry. 
\end{itemize}

\begin{figure}[htbp]
  \centering
  \begin{subfigure}[b]{0.45\linewidth}
    \includegraphics[width=\linewidth]{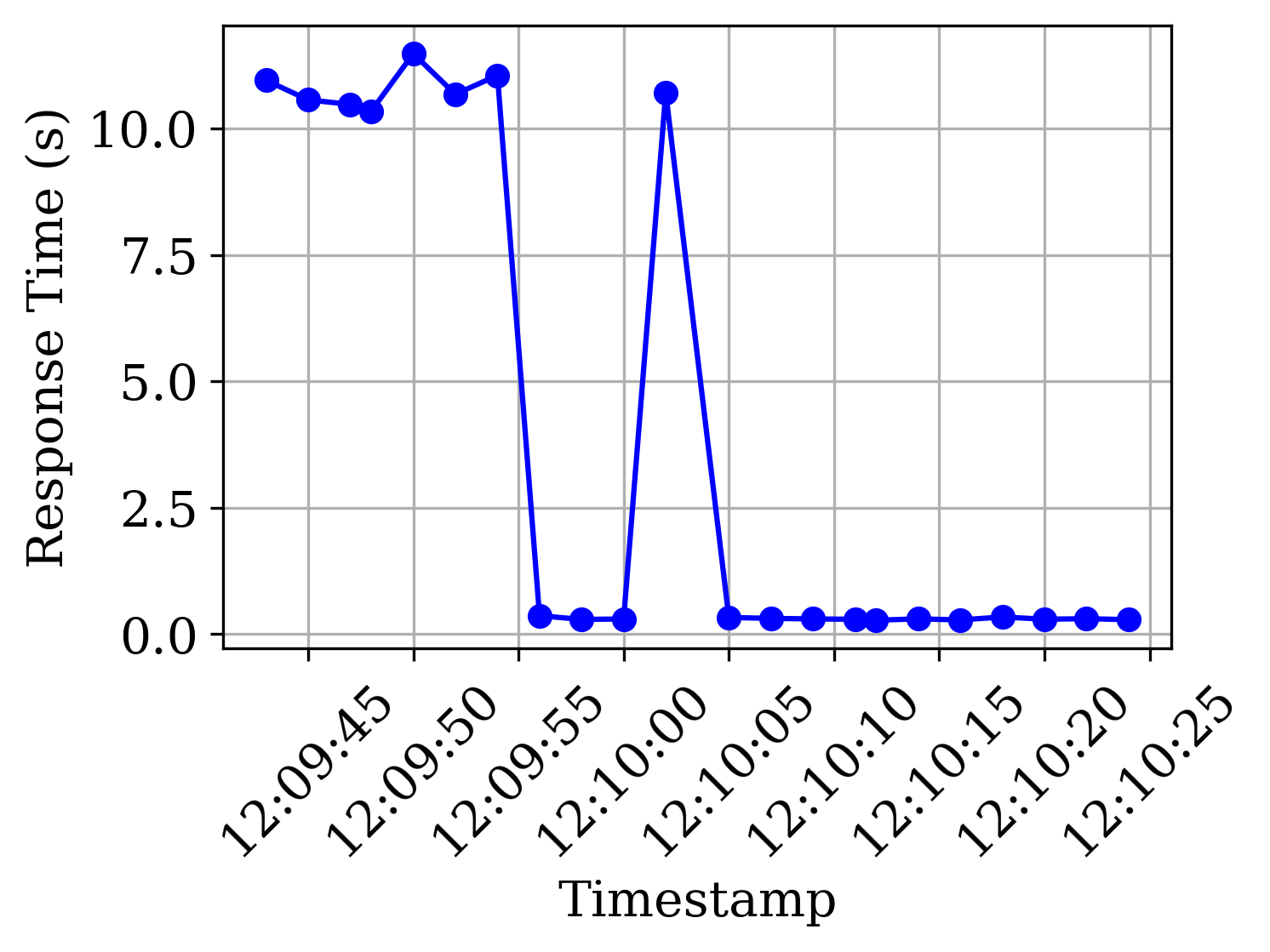}
    \caption{Cold start without warm container from start.}
    \label{fig:plot1}
  \end{subfigure}
  \hfill
  \begin{subfigure}[b]{0.45\linewidth}
    \includegraphics[width=\linewidth]{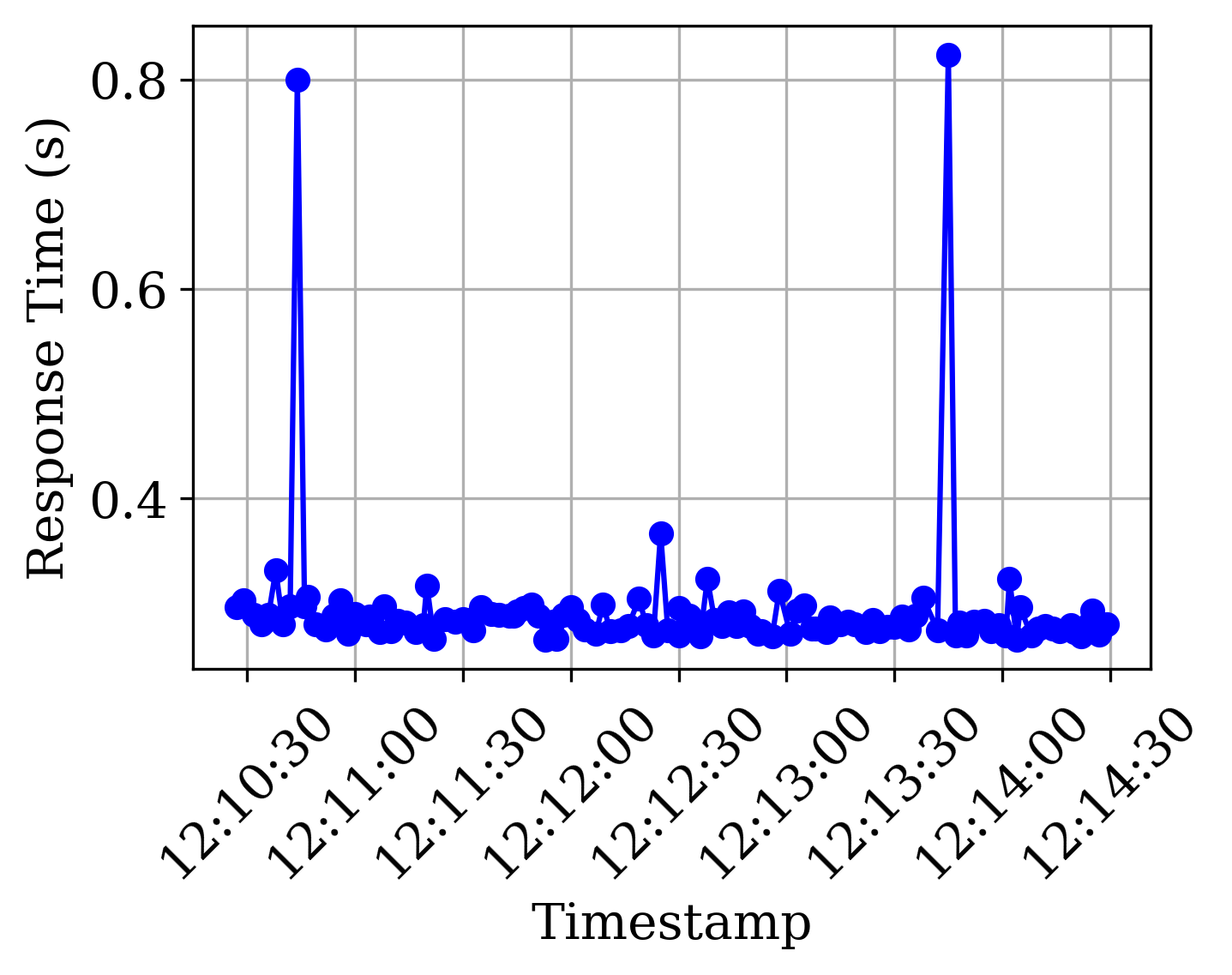}
    \caption{Normal situation.}
    \label{fig:plot2}
  \end{subfigure}

  \vspace{0.5em}  

  \begin{subfigure}[b]{0.45\linewidth}
    \includegraphics[width=\linewidth]{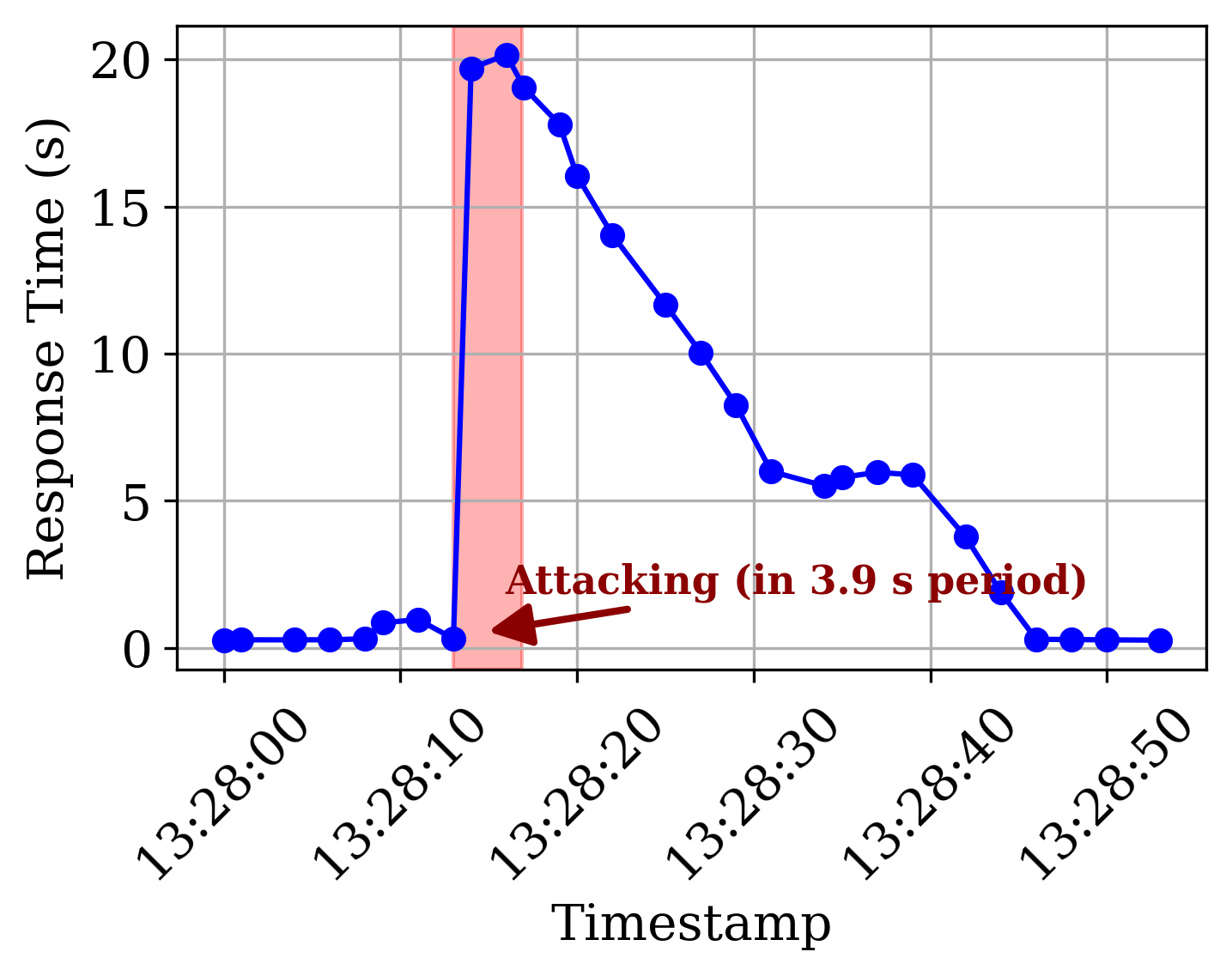}
     \caption{ Moderate-intensity DoS attack (59 req/s for 3.9 s).}
    \label{fig:plot3}
  \end{subfigure}
  \hfill
  \begin{subfigure}[b]{0.45\linewidth}
    \includegraphics[width=\linewidth]{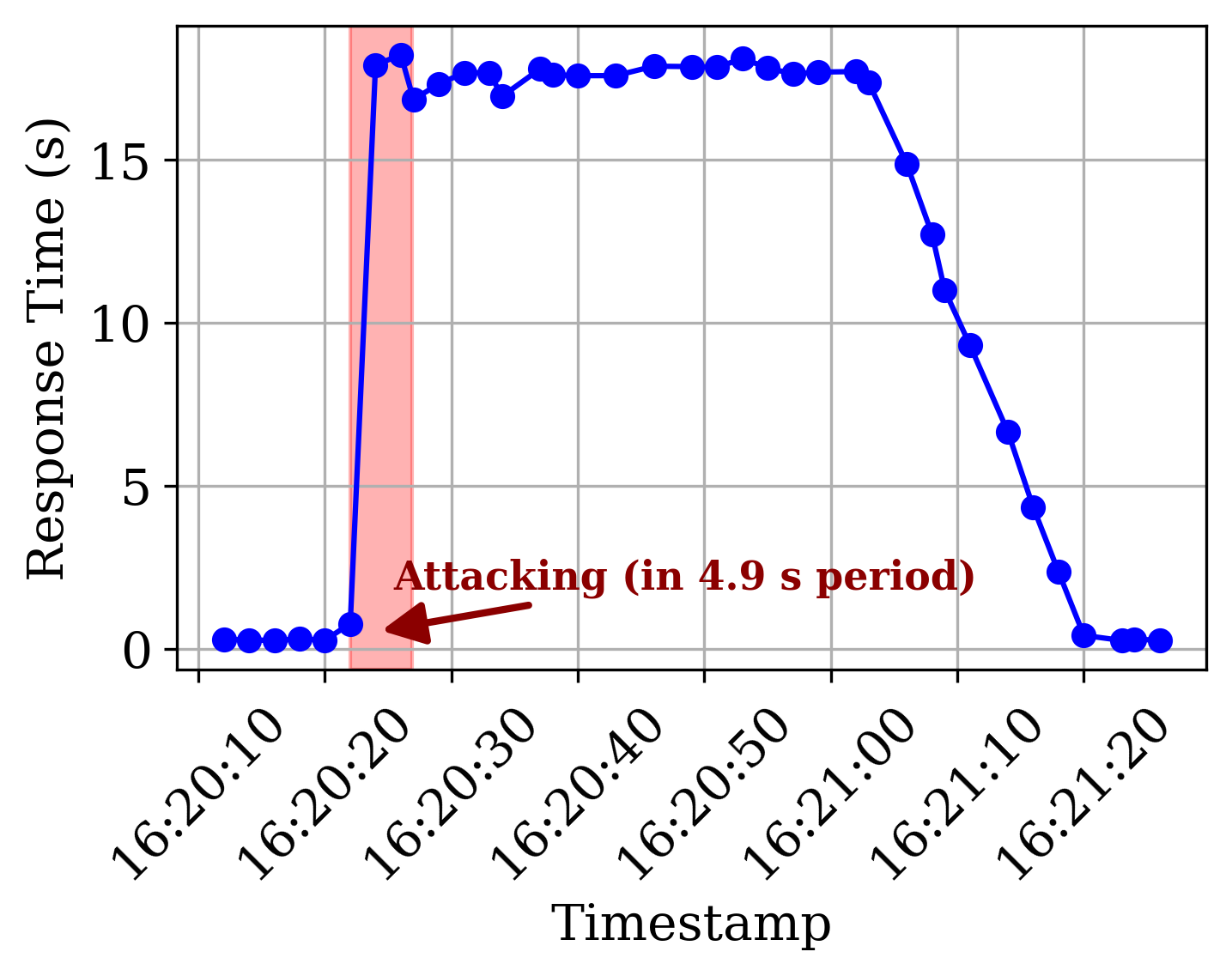}
    \caption{High-intensity DoS attack (203 req/s for 4.9 s).}
    \label{fig:plot4}
  \end{subfigure}

  \caption{End-to-end response time of functions - in (a) and (b) under normal conditions, and in (c) and (d) during attacks.}
  \label{fig:fourSubplots}
\end{figure}
\begin{figure}[htbp]
  \centering
  \begin{subfigure}[b]{0.45\linewidth}
    \includegraphics[width=\linewidth]{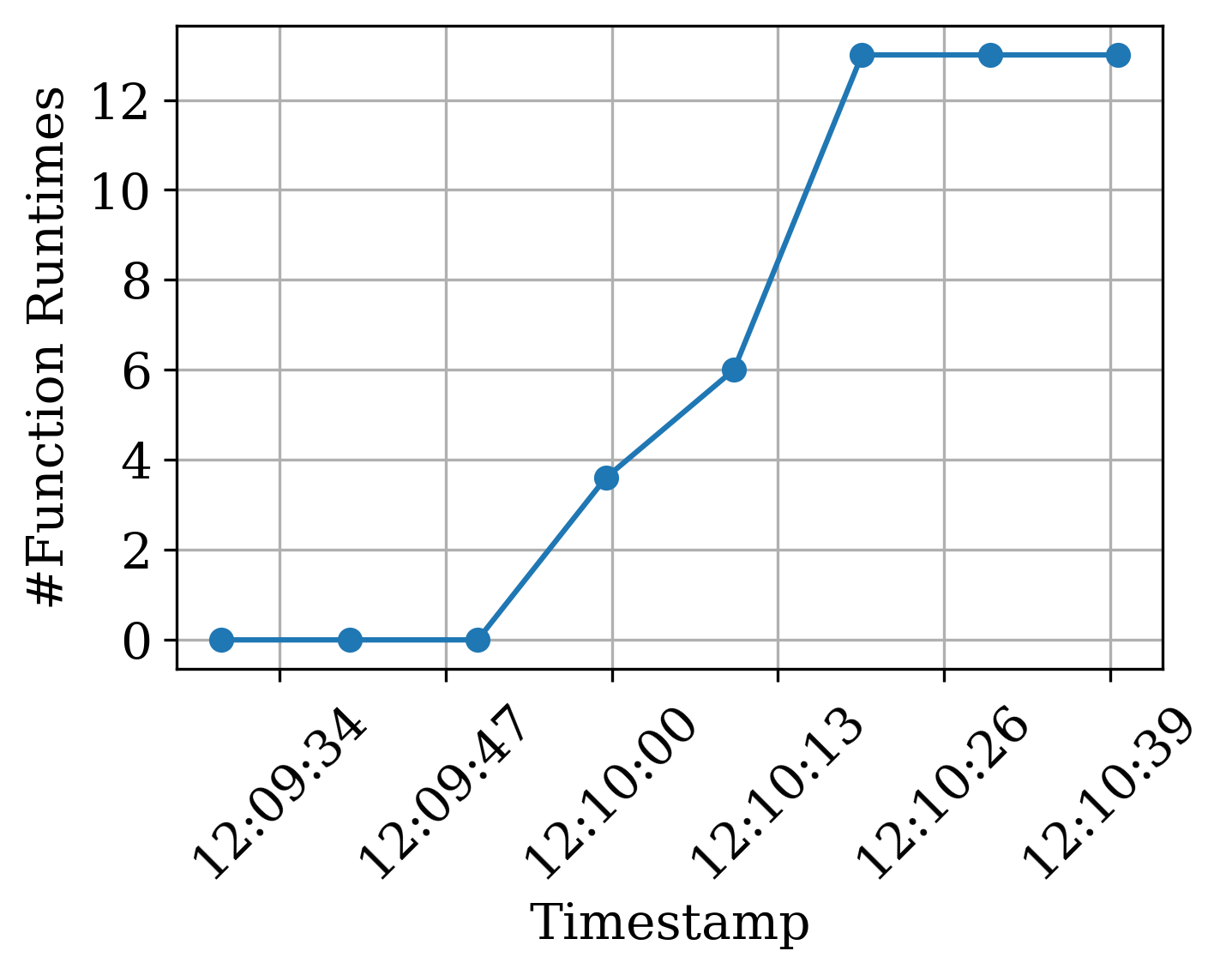}
    \caption{Cold start without warm container from start.}
    \label{fig:plot2a}
  \end{subfigure}
  \hfill
  \begin{subfigure}[b]{0.45\linewidth}
    \includegraphics[width=\linewidth]{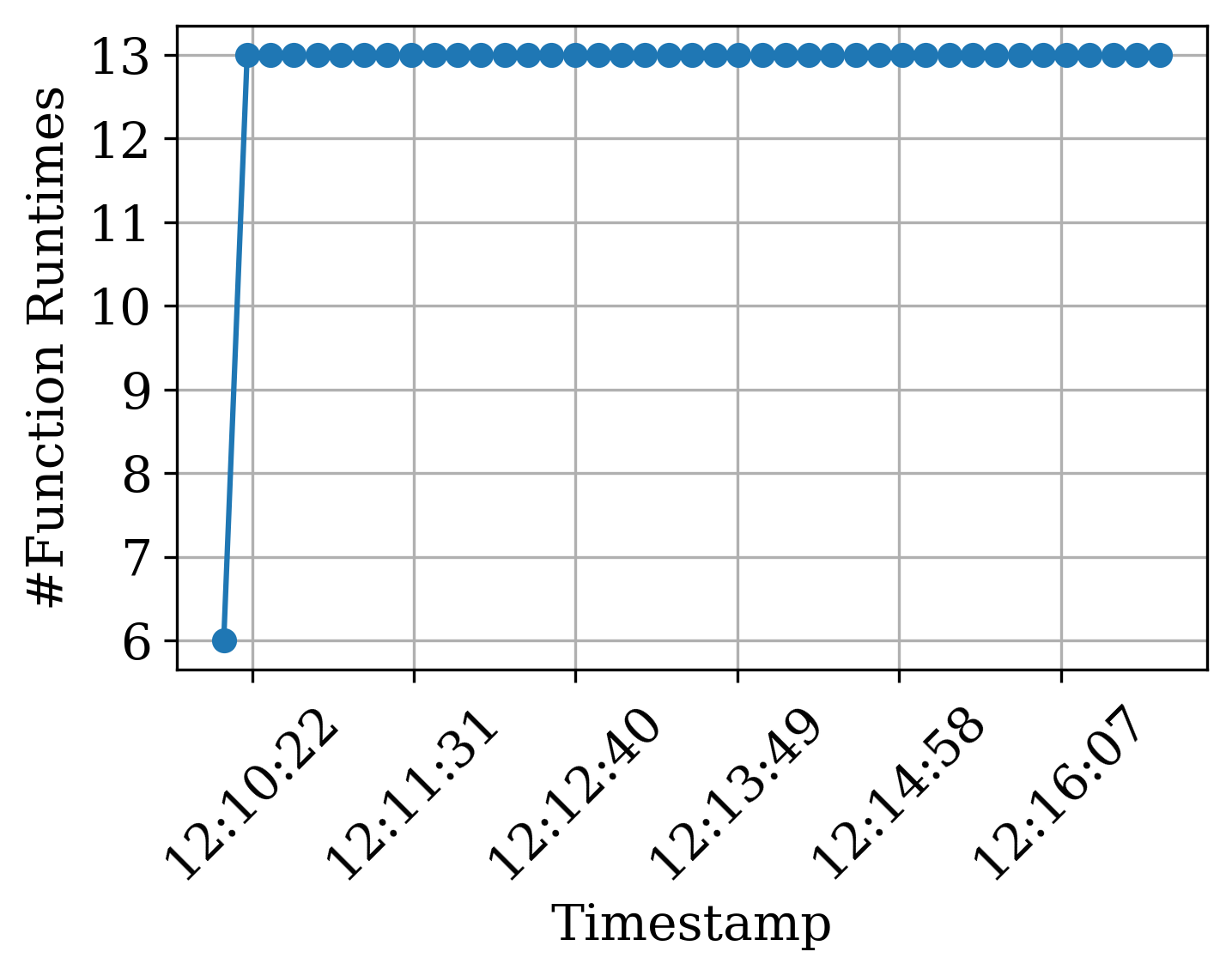}
    \caption{Normal situation.}
    \label{fig:plot2b}
  \end{subfigure}

  \vspace{0.5em}  

  \begin{subfigure}[b]{0.45\linewidth}
    \includegraphics[width=\linewidth]{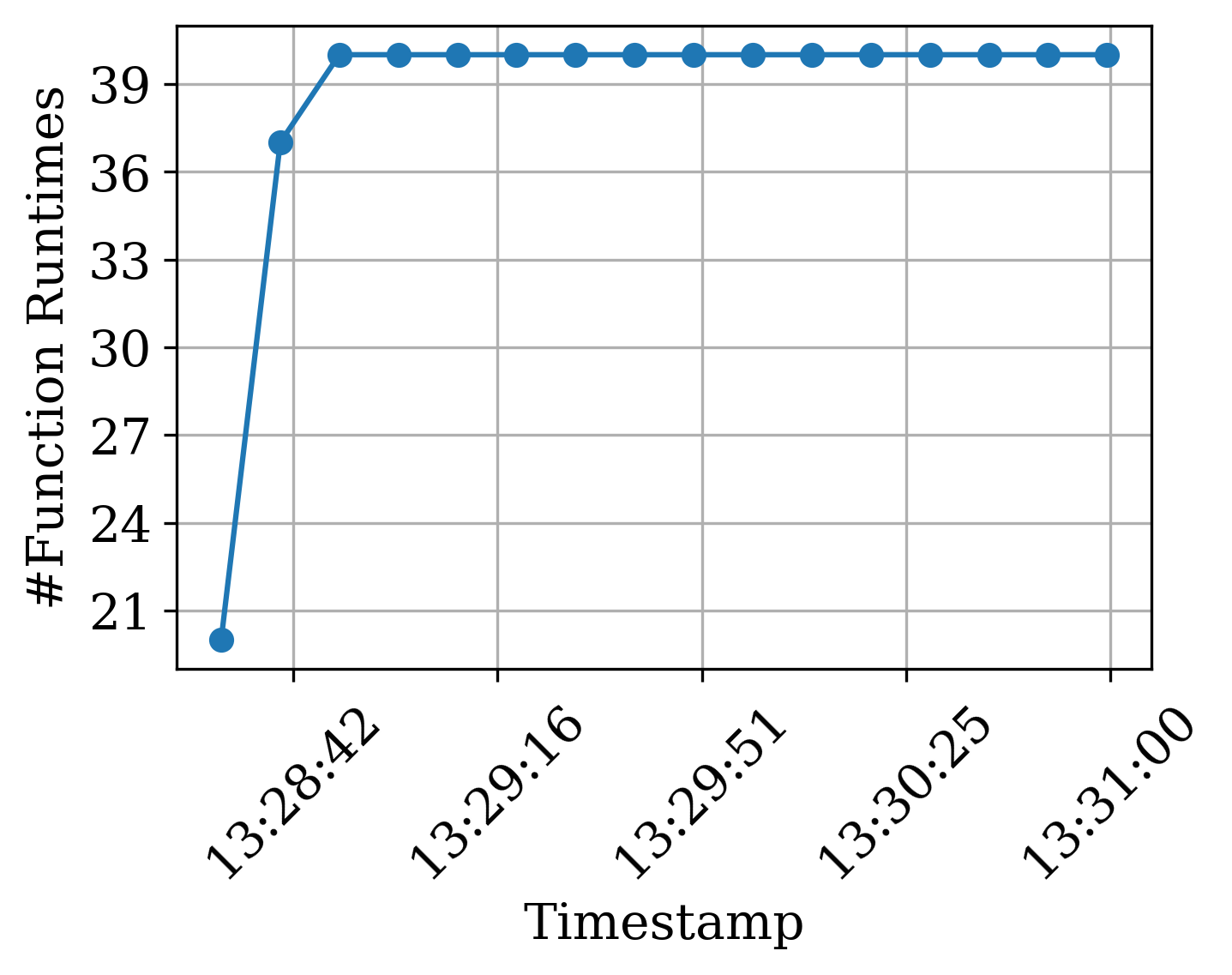}
    \caption{ Moderate-intensity DoS attack (59 req/s for 3.9 s).}
    \label{fig:plot2c}
  \end{subfigure}
  \hfill
  \begin{subfigure}[b]{0.45\linewidth}
    \includegraphics[width=\linewidth]{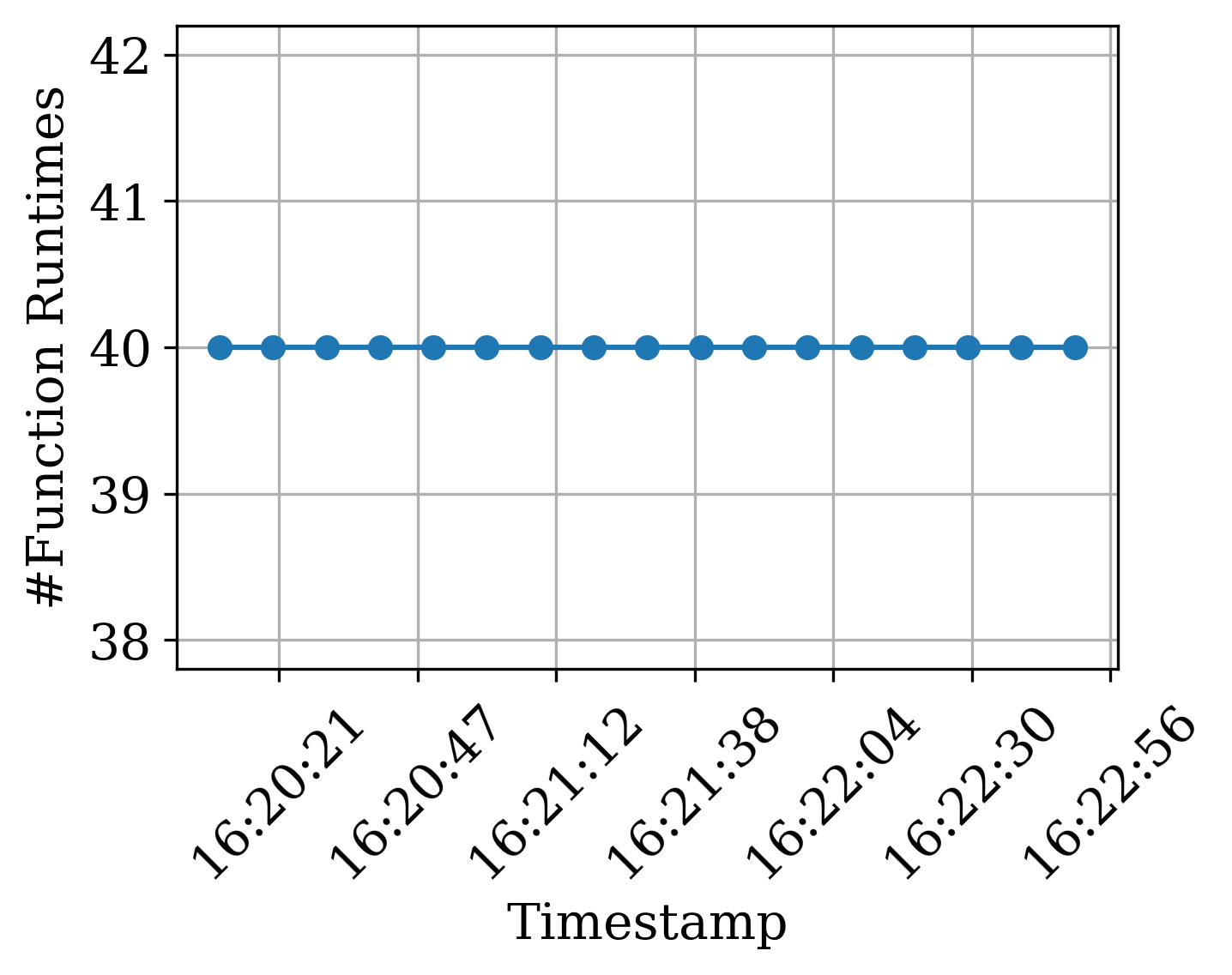}
    \caption{High-intensity DoS attack (203 req/s for 4.9 s).}
    \label{fig:plot2d}
  \end{subfigure}

\caption{Number of function instances over time, sampled every 10 seconds: (a) and (b) show normal conditions; (c) and (d) show behavior during attacks.}
  \label{fig:fplot2}
\end{figure}

\begin{figure}[htbp]
  \centering
  \begin{subfigure}[b]{0.45\linewidth}
    \includegraphics[width=\linewidth]{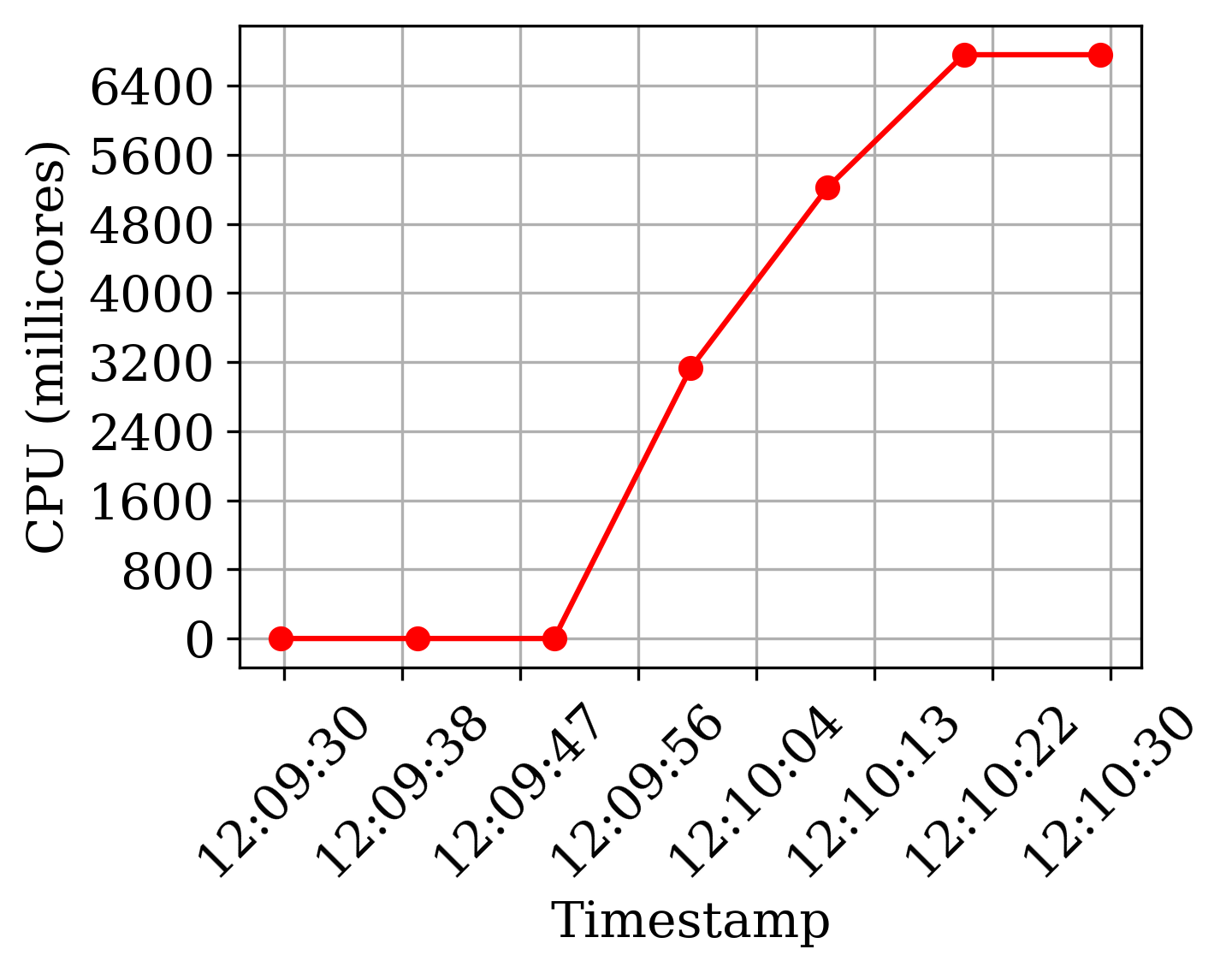}
    \caption{Cold start without warm container from start.}
    \label{fig:plot3a}
  \end{subfigure}
  \hfill
  \begin{subfigure}[b]{0.45\linewidth}
    \includegraphics[width=\linewidth]{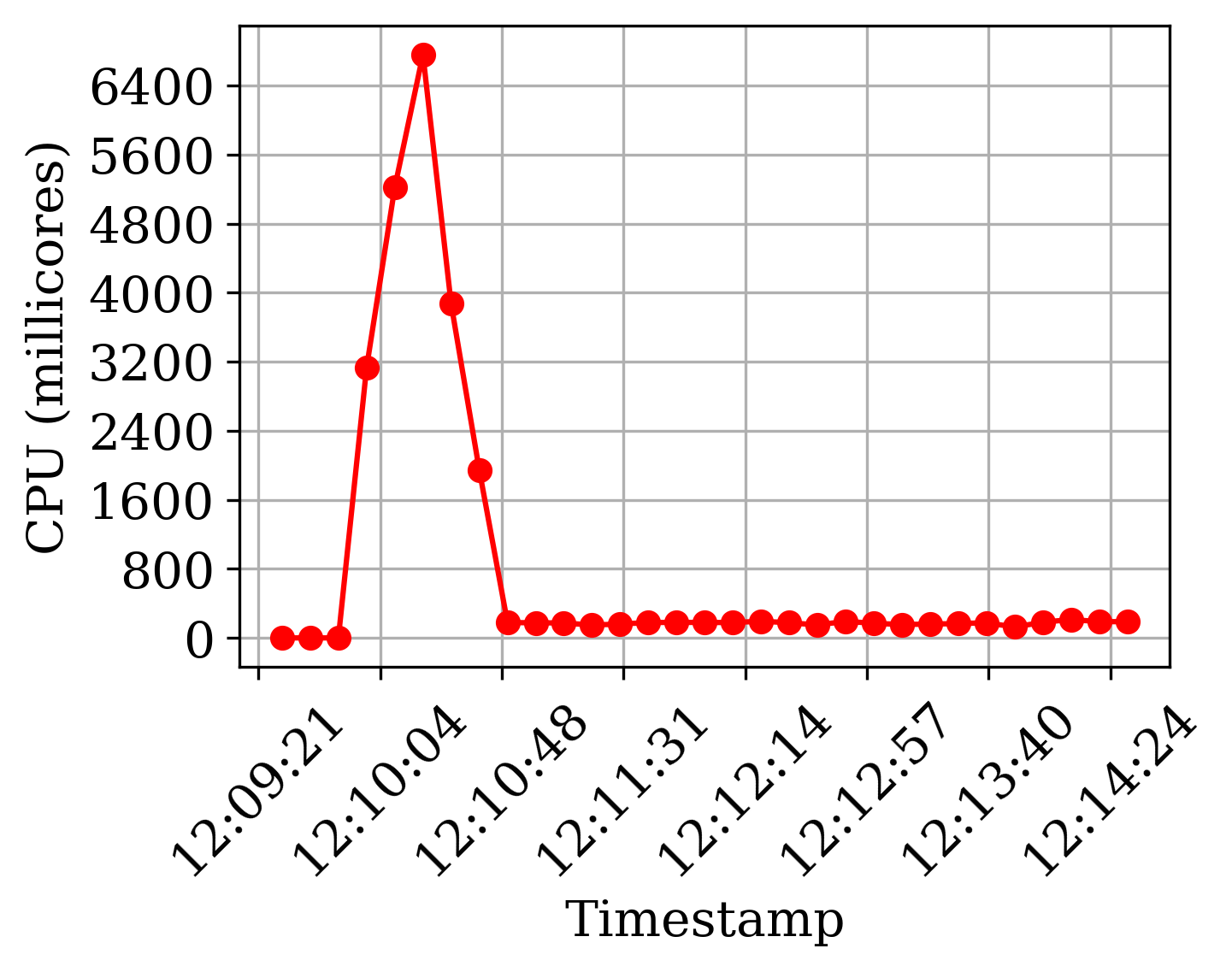}
    \caption{Normal situation.}
    \label{fig:plot3b}
  \end{subfigure}

  \vspace{0.5em}  

  \begin{subfigure}[b]{0.45\linewidth}
    \includegraphics[width=\linewidth]{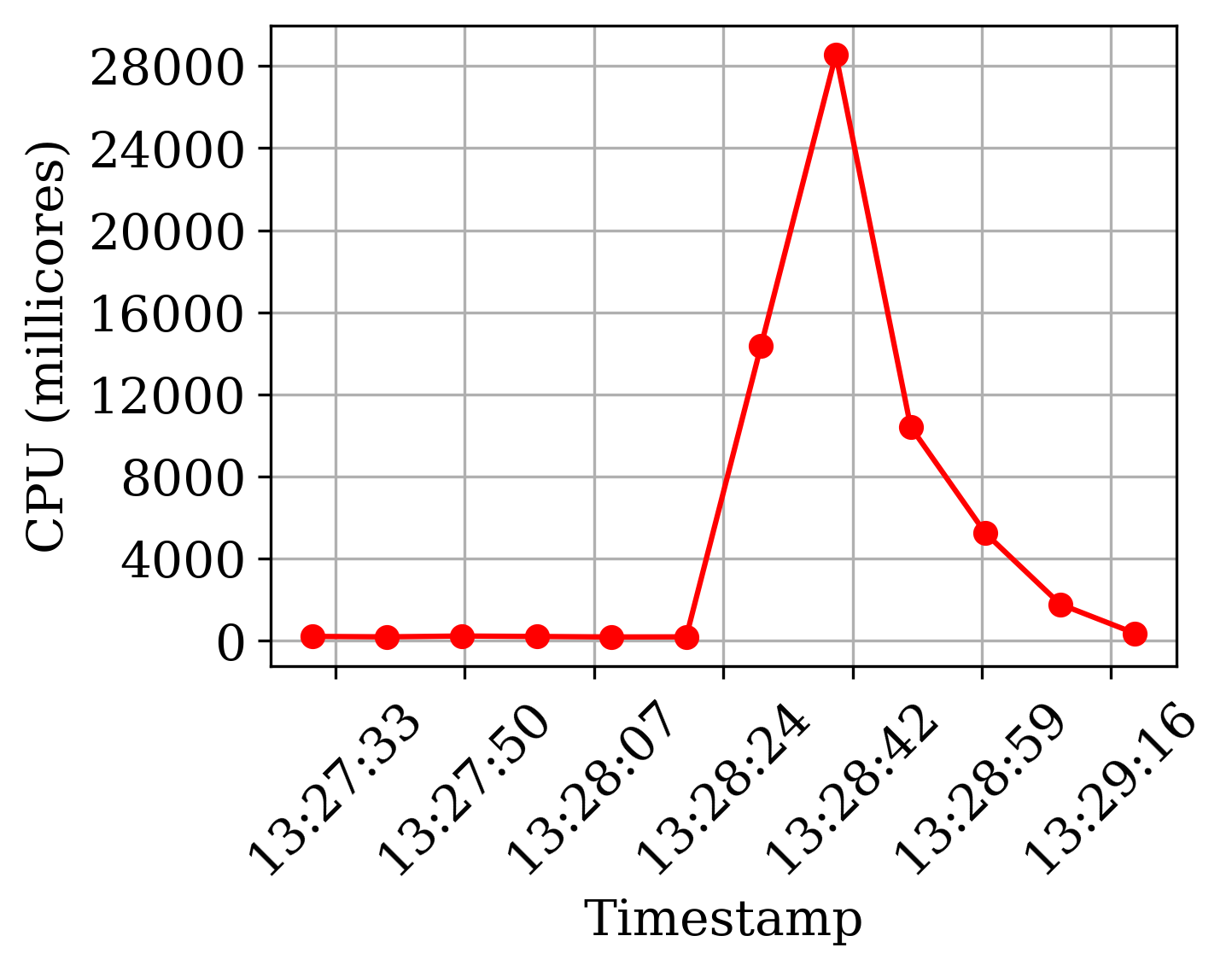}
    \caption{ Moderate-intensity DoS attack (59 req/s for 3.9 s).}
    \label{fig:plot3c}
  \end{subfigure}
  \hfill
  \begin{subfigure}[b]{0.45\linewidth}
    \includegraphics[width=\linewidth]{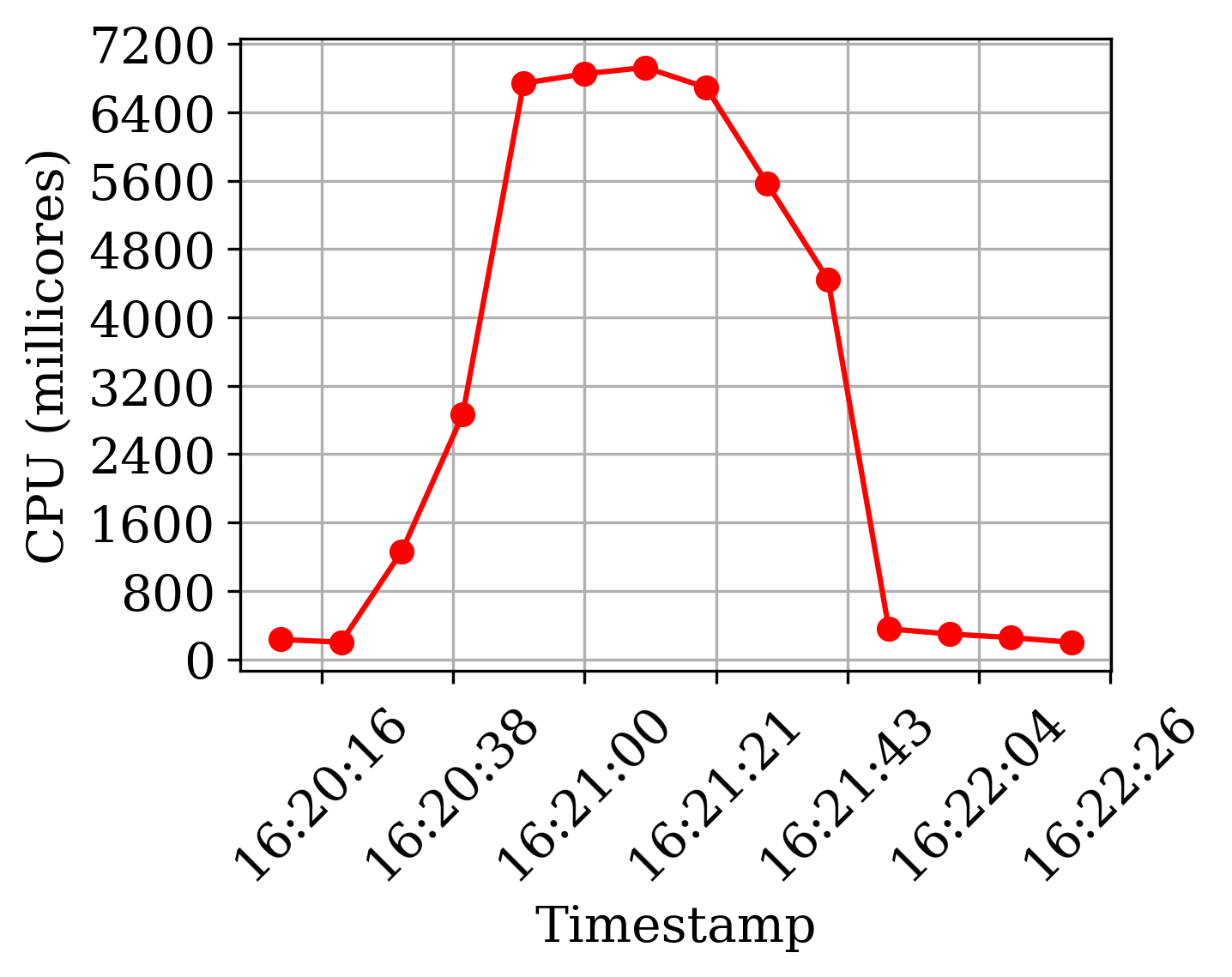}
    \caption{High-intensity DoS attack (203 req/s for 4.9 s).}
    \label{fig:plot3d}
  \end{subfigure}

\caption{Average CPU usage (millicores) per 10-second interval: (a) and (b) show normal conditions; (c) and (d) show behavior during attacks.}
  \label{fig:fplot3}
\end{figure}
\subsection{Challenges for anomaly detection in serverless environments}
\label{sec:cl}
To illustrate the challenges of anomaly detection in serverless environments, we present an empirical case study based on a latency-sensitive application deployed on the Apache OpenWhisk platform. The application implements a computer vision model, EfficientDet~\cite{tan2020efficientdet}, to support robotic systems in detecting objects of interest~\cite{nguyen2025tinykube}. In steady-state (warm) conditions, the function exhibits an average execution time of 280\,ms per request. However, when a cold start occurs, the additional initialization overhead inflates the end-to-end response time to approximately 10.5\,seconds. In serverless platforms, overall latency is further affected by queuing delays (e.g., captured by the \texttt{waitTime} metric in OpenWhisk), which represents the duration an invocation spends awaiting dispatch to an available container, particularly under conditions of high concurrency or resource contention.

To emulate adversarial behavior and evaluate detection robustness, we simulate a burst-style denial-of-service (DoS) attack. The attack injects short, high-intensity request bursts at random intervals, with randomized duration and intensity (i.e., number of concurrent invocations). These bursts are interleaved with idle periods during which only regular, benign function requests are issued. This setup mimics a stealthy attack scenario that aims to evade simple threshold-based detectors while still degrading system performance through cold start amplification and concurrency saturation.

Figure~\ref{fig:fourSubplots} presents the per-invocation response time across four scenarios: (a) cold start, where the system begins receiving requests without any warm containers; (b) normal operation, where the incoming request rate matches the number of available warm containers; (c) a moderate-intensity burst attack delivering 59 requests per second over 3.9 seconds; and (d) a high-intensity attack delivering 203 requests per second over 4.9 seconds.
Figures~\ref{fig:fplot2} and~\ref{fig:fplot3} further characterize the system’s behavior under the same four scenarios by showing, respectively, the number of warm containers (i.e., active function instances over time) and the average CPU usage in millicores.


The empirical observations from our experiments clearly highlight the unique difficulties in detecting and diagnosing anomalies within serverless environments.

First, the transient and stateless nature of serverless function instances makes it inherently difficult to establish stable baselines for ``normal” behavior -- a prerequisite for many anomaly detection techniques. In our cold start scenario (a), response times exceeded 10 seconds despite resource usage peaking at only 12 function instances and 6,400 CPU millicores. This elevated latency is not necessarily an “anomaly” in the traditional sense, but rather a fundamental characteristic of serverless scale-up. Distinguishing such expected high-latency startup phases from actual performance issues becomes challenging when instances are continuously being provisioned and deprovisioned.

Further, even under normal workload conditions (scenario b), CPU usage fluctuated rapidly, i.e., spiking to 6,400 millicores and quickly dropping to 200–300 millicores, while response times remained stable around 280\,ms, with only minor outliers. This \textit{volatility}, characteristic of \textit{ephemeral} and \textit{event-driven execution}, means that brief surges in resource usage or instance count may simply reflect normal platform elasticity rather than anomalies. Traditional detection methods that rely on consistent, long-lived signals often struggle to distinguish these legitimate bursts from true performance degradations.

Moreover, serverless functions react to external events without offering direct visibility into the orchestrator’s decisions or the underlying container lifecycle. This event-driven dynamism, combined with inherent observability limitations, complicates anomaly detection and root cause analysis. 
In our moderate attack scenario (c), a burst of 59 requests per second lasting 3.9 seconds caused response times to spike to nearly 20 seconds. Critically, although the attack was short-lived, its impact persisted: response times declined only gradually afterward. Simultaneously, the number of active function instances rose sharply from 20 to 40 and did not decrease after the burst subsided. CPU usage spiked to 28,000 millicores during the attack and then tapered off slowly, likely reflecting idle warm containers awaiting deallocation.
This illustrates how an overwhelmed event-driven system can enter a prolonged recovery phase that resembles normal autoscaling behavior but is in fact a lingering aftereffect of the attack. Without \textit{fine-grained insight} into the platform’s scaling mechanisms, it becomes difficult to determine when an attack begins or ends, and whether observed performance degradation reflects a malicious event or a benign workload shift.

In the high-intensity attack scenario (d), a burst of 203 requests per second over 4.9 seconds caused response times to remain above 15 seconds for more than a minute, demonstrating a broader and more persistent impact compared to the moderate attack. Function concurrency peaked at 40 instances, and CPU usage reached 7,200 millicores. However, the prolonged delay in response time recovery suggests deeper systemic strain, potentially due to queuing delays or resource contention within the platform.
Notably, the elevated number of function instances persisted well beyond the attack period. Yet, without \textit{visibility into per-instance workload or invocation queues}, it remains unclear whether these instances were actively serving requests or simply awaiting de-provisioning. The \textit{lack of transparency into the platform’s internal scheduling logic} makes it difficult to determine whether the platform is still recovering, failing to scale down, or encountering a hidden bottleneck. This ambiguity poses a major obstacle for timely and accurate anomaly detection.

Finally, response time, instance count, and CPU usage are typically collected from distinct observability layers. For example, API Gateway logs for response time, and platform-level metrics for resource usage. These metrics often differ in sampling intervals, aggregation granularity, and propagation delays. As a result, the causal relationships between high-level symptoms (e.g., latency spikes) and low-level platform behavior (e.g., instance scaling, cold starts, or CPU saturation) can be obscured or misaligned in time. This temporal and semantic disconnect complicates root cause analysis and can lead to delayed or inaccurate anomaly detection, particularly for short-lived or bursty events that escape coarse-grained aggregation windows.

Although our case study is grounded in a specific type of adversarial behavior, i.e., short burst-style denial-of-service attacks, many of the challenges it exposes are broadly applicable. The experiments serve not only to demonstrate the impact of such attacks but also to highlight fundamental limitations in current observability and anomaly detection approaches for serverless environments. In particular, the \textit{interplay between cold starts}, \textit{platform elasticity}, and \textit{lack of internal visibility} reveals that even intuitive, easily reproducible attacks can produce subtle, long-lasting performance degradation that evades conventional threshold-based detection. These insights underscore the need for more holistic, temporally aligned, and context-aware monitoring mechanisms capable of correlating high-level symptoms with low-level execution dynamics in serverless systems.


\section{A Vision for Next-Generation Detection Frameworks}
\label{sec:vision}
The challenges identified in our empirical study (Section~\ref{sec:threat}) underscore the critical need to fundamentally rethink anomaly detection for serverless platforms. Traditional techniques, built on assumptions of persistent services, stable baselines, and comprehensive full-stack observability, prove inadequate for the transient, opaque, and decentralized nature of serverless execution. Current approaches, which often rely on single-metric deviations or fixed thresholds, are particularly unreliable and prone to high false positive/negative rates, given the limited visibility into internal scheduling decisions and the highly abstracted nature of these platforms.

We envision a new class of detection frameworks that move beyond these limitations. Specifically, we propose two key ideas foundational to next-generation serverless anomaly detection: \textit{context-aware detection logic} and \textit{multi-source data fusion}. These concepts are detailed in the following subsections.

\begin{figure}[ht]
  \centering
  \includegraphics[width=\columnwidth]{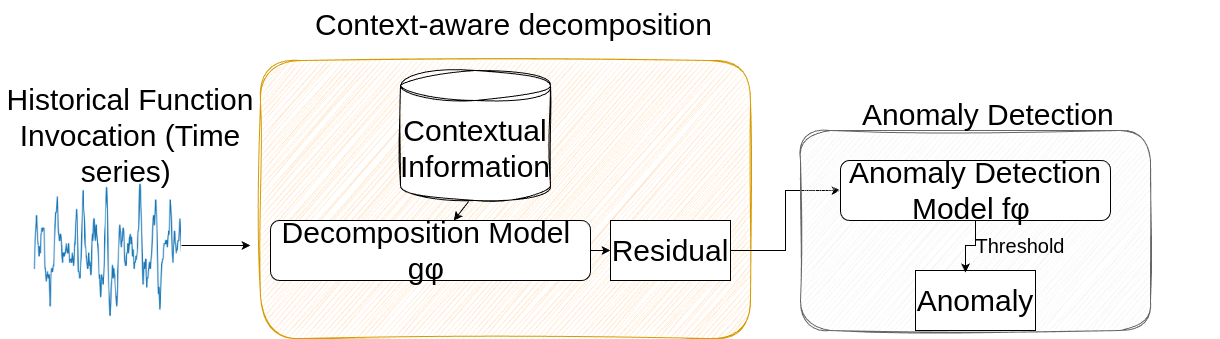}
  \caption{Concept of context-aware anomaly detection.}
  \label{fig:context_awre}
\end{figure}

\subsection{Context-aware detection logic} Effective anomaly detection in serverless environments fundamentally requires reasoning beyond raw metric deviations. Given the inherent ephemerality, statelessness, and bursty nature of function executions, establishing universally stable baselines for ``normal" behavior is often impractical; a metric value that is normal in one situation might be anomalous in another. Instead, detection must incorporate the immediate and broader execution context that dynamically shapes behavioral expectations. This includes transient factors like the presence of cold start phases, the characteristics of a specific workload burst (e.g., its volume and duration), or more stable patterns such as diurnal and weekly trends.

Inspired by state-of-the-art anomaly detection approaches in large-scale systems at Facebook, Microsoft, and Alibaba~\cite{zhang2022tfad, ren2019time, taylor2018forecasting}, we advocate for \textit{integrating sparse yet highly informative contextual variables} directly into the detection model. Such variables may include the precise time of day, the type of event that triggered the function invocation (e.g., HTTP request, AWS Simple Queue Service (SQS) message, or database change), the functional role of the invoked function (e.g., data ingestion, processing, or API endpoint), and the originating trigger source. These contextual elements help calibrate what constitutes expected metric values and define acceptable deviations more accurately.
Figure~\ref{fig:context_awre} illustrates the concept of context-aware anomaly detection.

For instance, consider a traffic-counting function deployed in a smart city. It would naturally experience morning and evening surges in invocation rates, along with corresponding latency or throughput peaks aligned with commuter flows. A spike in latency or throughput at 8:00 AM may therefore be entirely expected and reflect legitimate load, whereas the same pattern observed at midnight could indicate anomalous input traffic, backend slowdown, or even malicious trigger misuse. Furthermore, these expectations may vary spatially depending on the function’s deployment location. A sensor placed at a major intersection or highway on-ramp is likely to experience significantly higher loads during rush hours compared to one located in a quiet residential area or industrial zone.

To model such complex behaviors, raw metric streams (e.g., latency, invocation count) can be decomposed into trend, seasonal, and residual components, while simultaneously incorporating external contextual signals, such as endpoint location, time of day, and function type (for example, using attention-based neural architectures~\cite{nam2023context}). The residual component, which captures deviations not explained by trend or seasonality, serves as a refined and context-normalized input for anomaly scoring. Unlike traditional time-series models that operate solely on raw values, these approaches learn to weigh contextual cues when estimating expected behavior, thereby enabling more precise and robust anomaly detection.



\subsection{Multi-source data fusion}
\begin{figure}[ht]
  \centering
  \includegraphics[width=\columnwidth]{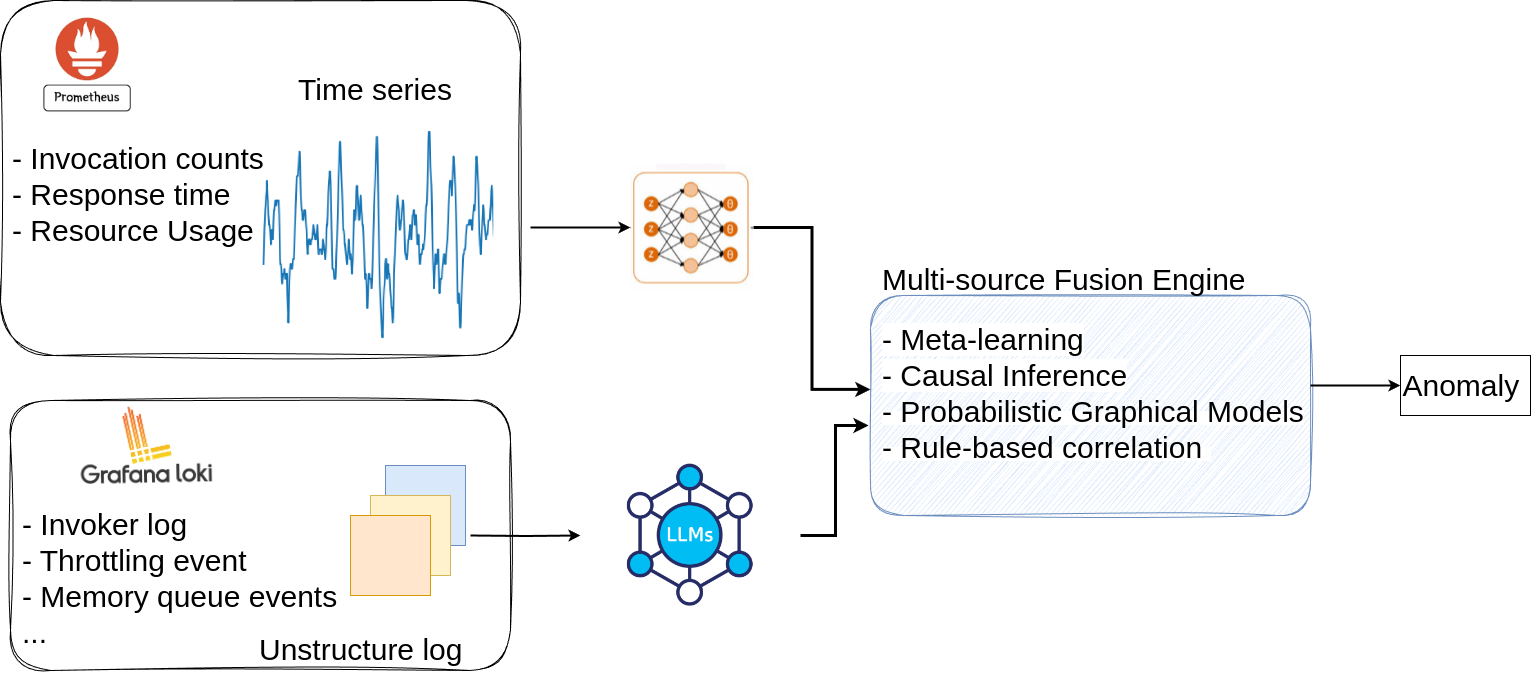}
  \caption{Concept of multi-source data fusion.}
  \label{fig:data_fusion}
\end{figure}
Robust anomaly detection in serverless environments requires reasoning across heterogeneous observability signals. Fusion mechanisms must integrate diverse data sources, such as response times, instance counts, cold start flags, reported resource usage, and event metadata, to capture coordinated anomalies that are invisible to any single metric in isolation. This is particularly critical in highly decoupled serverless architectures, where faults often propagate subtly across abstraction layers. Figure~\ref{fig:data_fusion} illustrates the conceptual architecture of a multi-source data fusion approach.

To support this vision, we propose a modality-specific detection pipeline in which distinct classes of AI models process structured and unstructured data sources. Structured telemetry, e.g.,  CPU and memory usage, invocation rates, and cold start events, can be continuously collected by observability systems like Prometheus and analyzed using traditional deep learning models (e.g., LSTMs \cite{Javad2021}, autoencoders \cite{jha2025llm}) to identify numerical deviations from historical baselines.

In parallel, unstructured logs from internal platform components, such as controller traces, invoker logs, and message queue events, provide a complementary, semantic view of system behavior. Generative AI models, particularly transformer-based language models~\cite{lee2023lanobert, chen2022bert}, are well-suited to this domain. When applied to logs ingested by platforms like Grafana Loki\footnote{\url{https://grafana.com/oss/loki/}}, these models can detect rare linguistic patterns, unusual event sequences, or contextual shifts indicative of anomalies that are otherwise undetectable via metrics alone.

The outputs of these domain-specialized models (ranging from anomaly scores to learned embeddings) are then combined via a high-level fusion engine. This engine, potentially based on meta-learning~\cite{fang2020meta}, causal inference~\cite{bareinboim2016causal}, or probabilistic graphical models~\cite{makarenko2009decentralised, koller2009probabilistic}, integrates multi-modal evidence to infer system-wide anomalies. Such fusion is essential for surfacing complex issues like degraded scheduling efficiency, noisy neighbor interference, or lingering post-attack effects -- cases where neither metrics nor logs alone provide a complete picture.

We emphasize that achieving this level of holistic detection depends on deep platform instrumentation and privileged access to both telemetry and internal logs, i.e., capabilities typically reserved for cloud providers or administrators of custom serverless deployments. As such, this approach is most applicable to platform-facing detection tools aimed at improving resilience, transparency, and operator trust in large-scale serverless systems.

Beyond these core capabilities, the practical deployment of next-generation serverless anomaly detection frameworks hinges on adherence to several key design principles. First, detection mechanisms must be \textit{lightweight and capable of real-time inference}, as serverless platforms are characterized by high invocation volume and short function lifespans. This necessitates computationally efficient models, e.g, statistical or learning-based, that can operate under strict latency and resource constraints.

Second, \textit{privacy-preserving design} is essential, particularly as serverless applications increasingly process sensitive data. Detection systems should minimize deep payload inspection and support decentralized analysis to reduce data exposure across tenants or administrative domains.

Finally, with the rise of distributed serverless deployments spanning heterogeneous edge and cloud nodes, edge–cloud split awareness becomes critical. This entails partitioning the detection logic: lightweight components can perform local filtering or statistical sketching at the edge, while more intensive correlation and inference are offloaded to the cloud. 
Together, these principles define what it means for anomaly detection to be serverless-native.
\section{Research Opportunities and Open Questions}
\label{sec:opp}
Our vision for next-generation serverless anomaly detection highlights a significant gap between current methodologies and the complex operational realities of modern FaaS platforms. While we have outlined key design principles and proposed promising directions, numerous fundamental research opportunities and open questions remain. Addressing these will be critical for advancing the field towards robust, scalable, and truly resilient serverless systems.

\subsection{Data generation, labeling, and benchmarking}
A key prerequisite for advancing anomaly detection in serverless environments is the availability of realistic, diverse, and accurately labeled datasets. Several important questions raised regarding realistic anomaly generation and injection: How can we effectively generate or inject diverse types of anomalies (e.g., subtle resource contention, sophisticated adversarial exploits, cascading failures) into serverless platforms without compromising live systems? Developing realistic synthetic anomaly injection frameworks that faithfully mimic real-world threat actors and operational issues is a significant open problem. Further, given the ephemeral nature and distributed complexity of serverless, acquiring accurate ``ground truth" labels for anomalies is exceedingly difficult. Research is needed into semi-supervised or unsupervised learning techniques, active learning strategies for expert-guided labeling, and methods for weakly labeling anomalies based on platform-level alerts or post-mortem incident reports.
At present, the absence of standardized benchmarks and public datasets hinders direct comparison and evaluation of proposed detection frameworks. How can we design and implement comprehensive benchmarking frameworks that account for serverless-specific attributes like cold starts, bursty workloads, multi-tenancy, and diverse deployment scenarios (e.g., edge vs. cloud)? This includes defining appropriate metrics for evaluation that go beyond traditional precision/recall to quantify operational impact.

\subsection{Advanced inference and interpretability}
Our proposed context-aware and multi-source data fusion approaches rely on sophisticated AI techniques, which introduce their own set of challenges. As detection models become more complex (e.g., deep learning, generative AI), their decision-making often becomes opaque. Operators need actionable explanations for why an anomaly was flagged to facilitate timely investigation and mitigation. How can we develop Explainable AI (XAI) techniques~\cite{dwivedi2023explainable} that provide interpretable insights, linking model decisions back to specific log patterns, metric deviations, or contextual factors, while accommodating the transient nature of function execution?

In parallel, enabling real-time inference over fused multimodal data streams, including structured metrics, unstructured logs, and distributed traces, remains an open technical challenge. Developing low-latency, scalable fusion engines that can represent and combine heterogeneous inputs in a unified manner requires advances in cross-modal representation learning and robust data fusion architectures. These systems must operate under high throughput, tolerate missing or noisy data, and adapt to evolving workloads without retraining from scratch. Meeting these requirements is essential for deploying next-generation anomaly detection in large-scale, production-grade serverless environments.

Finally, anomalies require not just detection, but also automated or semi-automated mitigation. How can detection frameworks seamlessly integrate with platform control planes to trigger adaptive responses (e.g., scaling adjustments, function isolation, dynamic rate limits) that are tailored to the anomaly's root cause and impact, minimizing disruption and ensuring system resilience? This includes strategies for graceful degradation during platform-wide stress events.

\subsection{Advanced observability models for diagnosing anomalies}
Observability models for diagnosing anomalies in serverless systems must go beyond traditional metrics and logging to address the unique challenges of ephemeral, stateless, and highly distributed environments. These models should be context-aware, enabling granular visibility into individual function behaviors and their roles within diverse workflows and systems. They may be trace-driven, capable of capturing end-to-end execution paths to uncover causal relationships across asynchronous function invocations. Context enrichment is essential for interpreting anomalies accurately, incorporating runtime metadata such as input parameters, dependency states. Crucially, these models should support semantic correctness validation, ensuring that outputs are not only generated but also logically accurate and align application's intent.  


\section{Realization Challenges}
\label{sec:challenges}

While this paper outlines a vision for anomaly detection tailored to the unique nature of serverless computing, several key challenges must be addressed to translate this vision into practice.

First, realizing real-time, lightweight detection frameworks depends on access to fine-grained telemetry data, such as per-invocation traces, inter-function dependencies, and event lineage. However, current serverless platforms often limit observability due to abstraction, multi-tenancy, and proprietary constraints. This lack of transparency makes both anomaly attribution and root-cause analysis significantly harder.

Second, integrating novel anomaly detection systems into real-world serverless pipelines presents a distinct set of deployment challenges. These include accounting for the stringent cost constraints inherent in serverless models, ensuring robust resource isolation for detection components themselves, guaranteeing privacy preservation when analyzing sensitive data, and maintaining compatibility across diverse and evolving cloud provider environments. Critically, ensuring that the detection system itself does not introduce undue latency or significant overhead to the serverless application's execution path or resource consumption is paramount for its practical viability.

Finally, the serverless threat landscape is still evolving. Emerging attack vectors and usage patterns may render existing detection strategies obsolete or insufficient. As such, any future detection architecture must be inherently adaptable, capable of learning from new behaviors, reacting to unseen anomalies, and evolving with the platform ecosystem.

These substantial challenges, far from diminishing our proposed vision's relevance, instead underscore its urgency. They illuminate a critical and promising research agenda: developing resilient, adaptive, and cost-effective anomaly detection techniques specifically designed for the dynamic, distributed nature of serverless computing.

\section{Conclusion}
\label{sec:conclus}
Serverless computing is rapidly transforming cloud application development, offering immense scalability and simplified deployment through automatic scaling and fine-grained billing. In these dynamic, event-driven environments, even brief disruptions, such as latency spikes, execution failures, or unexpected cost anomalies, can severely impact user experience and operational efficiency. As a result, anomaly detection becomes essential for continuously monitoring system behavior, identifying deviations from expected norms, and enabling timely mitigation of emerging issues.

However, the intrinsic characteristics of serverless platforms pose significant challenges to effective anomaly detection. In this vision paper, we meticulously identified these specific challenges, including the absence of persistent contexts, abstracted runtimes, event correlation difficulties, and monitoring granularity gaps. We also explored the full spectrum of operational vulnerabilities and novel adversarial threats, from Denial of Service and Denial-of-Wallet attacks to more sophisticated exploits such as cold start amplification.

Building on these insights, we articulated a compelling vision for next-generation anomaly detection frameworks. Our proposed research agenda centers on techniques that are context-aware, leverage multi-source data fusion, operate in real-time, prioritize privacy, and adapt to edge–cloud deployments. By laying this foundational roadmap, we aim to drive the development of robust, anomaly-resilient serverless ecosystems, crucial for unlocking the full potential of this transformative cloud paradigm.
\section*{Acknowledgement} 
This work was supported by the European Commission through the Horizon Europe project SovereignEdge. COGNIT (grant no. 101092711). Additional support was provided by the Wallenberg AI, Autonomous Systems and Software Program (WASP) funded by Knut and Alice Wallenberg Foundation.

\bibliographystyle{ieeetr}
\balance
\bibliography{acmart}

\end{document}